\documentclass[aps,pra,reprint]{revtex4-1}
\usepackage{graphicx}
\usepackage{amsmath,amssymb}
\usepackage{fullpage}
\usepackage{siunitx}
\usepackage{todonotes}
\usepackage{hyperref}

\begin{document}

\newcommand{\inlinecode}{\texttt}

\title{Serial Electron Diffraction Data Processing with \emph{diffractem} and \emph{CrystFEL}}
    
\author{Robert Bücker}
\affiliation{Max Planck Institute for the Structure and Dynamics of Matter, Center for Free-Electron Laser Science, Luruper Chaussee 149, 22761 Hamburg, Germany}
\affiliation{Centre for Structural Systems Biology, Department of Chemistry, University of Hamburg, Notkestraße 85, 22607 Hamburg, Germany}
\email{robert.buecker@mpsd.mpg.de}
\author{Pascal Hogan-Lamarre\footnotemark[1]}
\affiliation{Max Planck Institute for the Structure and Dynamics of Matter, Center for Free-Electron Laser Science, Luruper Chaussee 149, 22761 Hamburg, Germany}
\affiliation{Departments of Physics and Chemistry, University of Toronto, 80 St.\ George Street, Toronto, ON M5S 3H6, Canada}
\author{R. J. Dwayne Miller}
\affiliation{Departments of Physics and Chemistry, University of Toronto, 80 St.\ George Street, Toronto, ON M5S 3H6, Canada}

\begin{abstract}
    Serial electron diffraction (SerialED) is an emerging technique, which applies the snapshot data-collection mode of serial X-ray crystallography to three-dimensional electron diffraction (3D~ED), forgoing the conventional rotation method. 
    Similarly to serial X-ray crystallography, this approach leads to almost complete absence of radiation damage effects even for the most sensitive samples, and allows for a high level of automation.
    However, SerialED also necessitates new techniques of data processing, which combine existing pipelines for rotation electron diffraction and serial X-ray crystallography with some more particular solutions for challenges arising in SerialED specifically.
    Here, we introduce our analysis pipeline for SerialED data, and its implementation using the \emph{CrystFEL} and \emph{diffractem} program packages.
    Detailed examples are provided in extensive supplementary code.
\end{abstract}

\maketitle

\section{Introduction}

Transmission electron microscopy as a tool for both material and life science has recently seen revolutionary developments, driven by new types of electron detectors, computational data analysis, automation, and sample preparation.
Concomitantly, statistics from the Protein Data Bank (PDB) and the Electron Microscopy Data Bank (EMDB) show a clear increase in the number of protein structures that are recovered through electron-based techniques.
Indeed, cryo-electron microscopy (CryoEM) produces the majority of the protein structures in the 3.5-5~\si{\angstrom} resolution range that are being released nowadays.
The predominant CryoEM techniques comprise single-particle analysis and tomography, the former being especially suitable for elucidating the structure of proteins and larger complexes at near-atomic resolution, whereas the latter allows to image larger, inhomogeneous structures, up to entire cells.
However, single-particle analysis is limited in its scope to molecules of weight above \(\approx \SI{40}{kDa}\), as the signal-to-noise ratio of such small particles in electron micrographs is not sufficient for computational alignment~\cite{Henderson1995,Glaeser2019}, and despite recent progress in CryoEM~\cite{Nakane2020,Yip2020}, X-ray crystallography is still clearly predominant for routine structure determination at the atomic resolution scale.
Diffractive electron techniques such as crystallography of monolayers of proteins (2D crystallography) led to seminal results~\cite{Henderson1975,Henderson1990,Gonen2005}, but ultimately remained limited in scope as preparation of suitable two-dimensional crystals is often prohibitively difficult.

On the other hand, there have been successful implementations of three-dimensional electron diffraction (3D ED/MicroED) techniques, where three-dimensional, sub-micron-sized crystals are used, in analogy to X-ray crystallography~\cite{Gemmi2019, Nannenga2019}.
As the interaction of electrons with matter is increased by up to six orders of magnitude with respect to X-ray photons, sizable signals can be obtained from even tiny crystals.
This, combined with the high dose efficiency of electrons, that is, a favorable ratio of elastic to inelastic events and small energy release during inelastic events, and the signal amplification afforded by diffraction-mode acquisition~\cite{Clabbers2018a} makes 3D ED especially appealing for materials which form only small and radiation-sensitive crystals.
The potential of 3D ED techniques have first been realized in material science~\cite{Kolb2007,Zhang2010}.
Excellent results could be obtained for radiation-sensitive nanocrystalline materials such as zeolites~\cite{Su2014}, or covalent- and metal-organic frameworks~\cite{Zhang2013}, which often evade X-ray structure determination.
Soon after, 3D ED has been introduced into life science (there mostly known as MicroED)~\cite{Shi2013,Nederlof2013,Nannenga2014a}, where high-resolution structures of small proteins, peptides and pharmaceuticals can now routinely be solved~\cite{Nannenga2019}.

Most of the 3D ED/MicroED work has so far been performed by rotating the crystal in the electron beam in various ways~\cite{Gemmi2019a}, in analogy to goniometer-based X-ray single-crystal diffraction.
More recently, \emph{serial} electron diffraction (SerialED) has been introduced~\cite{Smeets2018,Buecker2020}, where, in analogy to synchrotron- and free-electron laser-based techniques~\cite{Chapman2019,Gati2014,Stellato2014}, a large ensemble of nanocrystals is employed, each of which only a single diffraction pattern is taken from.
While this data collection scheme has important advantages over rotation methods, it requires a different approach to data processing, specifically in the data-reduction steps of a crystallographic pipeline, from raw data to estimated Bragg reflection intensities.

In this paper, we discuss our pipeline for SerialED data processing.
The paper is structured as follows:
In Section~\ref{sec:serialed}, we briefly recapitulate the concept of SerialED and its implementation in our laboratory, as described in~\cite{Buecker2020}.
Next, in Section~\ref{sec:pipeline}, we discuss the general data processing pipeline, illustrated by examples from a typical data set.
Section~\ref{sec:diffractem} introduces our program package \emph{diffractem} and outlines it usage for the pipeline described in Section~\ref{sec:pipeline}.
Finally, Section~\ref{sec:discussion} reviews various specific aspects and potential issues of our approach, and future directions of further development.

\section{Serial Electron Diffraction: concept and data collection}
\label{sec:serialed}

While rotation crystallography, whether using electrons or X-rays, can yield high-quality crystallographic data from nanometric crystals, an inherent limitation is the accumulation of radiation damage during rotation data collection~\cite{Hattne2018}, prohibiting acquisition of damage-minimized data.
On the other hand, damage accumulation is evaded in serial crystallography, where each crystal is exposed once, using femtosecond X-ray pulses at extreme intensities that record diffraction data before Coulomb explosion~\cite{Chapman2011}, or X-ray/electron pulses at lower intensity below a critical dose threshold, which can yield equivalent results~\cite{Mehrabi2020,Buecker2020}.

\begin{figure}
    \centering
    \includegraphics[width=0.95\columnwidth]{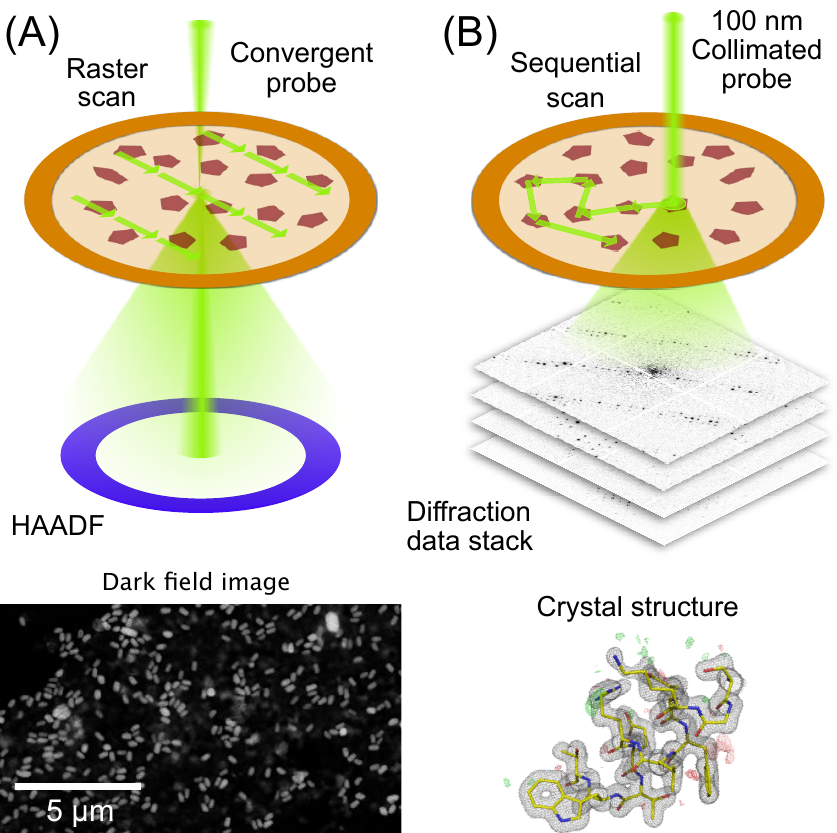}
    \caption{Principle of STEM-based SerialED. 
    (A) A low-resolution, low-dose STEM image is taken over a large region on a TEM grid.
    Signal is generated using the high-angle dark field (HAADF) detector.
    (B) After the crystals have been identified in the STEM image, the beam is sequentially steered to each autmatically found crystal.
    A fast detector records the diffraction patterns in a synchronized way.
    From the diffraction data, the crystal structure is solved.}
    \label{fig:serialed}
\end{figure}

To automate the process of collection of diffraction data from thousands of crystals randomly dispersed on an electron microscope grid, serial electron diffraction (SerialED) leverages the ability of electron microscopes to map out their locations, using conventional~\cite{Smeets2018} or scanning~\cite{Buecker2020} TEM imaging (Figure~\ref{fig:serialed}~A).
Crystals are automatically identified in the map image, and the electron beam is steered sequentially to the found crystals, where diffraction patterns are taken (Figure~\ref{fig:serialed}~B). 
The process can then be repeated in many regions of a sample grid, each typically tens of \si{\micro \metre} across.
This approach adds a high level of automation to the advantages of SerialED, requiring little specific skill on the user's part for operation. 
In~\cite{Buecker2020}, SerialED was furthermore combined with a dose fractionation scheme as known from single-particle electron microscopy, which allows to obtain damage-minimized data as described above, without the need for prior information about the sample or exact calibrations.

Despite these advantages, with respect to rotation techniques, SerialED poses new challenges with regards to data analysis, specifically pertaining to the steps of data reduction from raw diffraction patterns to merged Bragg spot intensities.
In this article, we discuss the processing of SerialED data sets using \emph{CrystFEL}~\cite{White2012} and \emph{diffractem}, a new library specifically developed for SerialED.

\section{Processing Method for Serial Electron Diffraction}
\label{sec:pipeline}

\begin{figure*}
    \centering
    \includegraphics[width=0.95\textwidth]{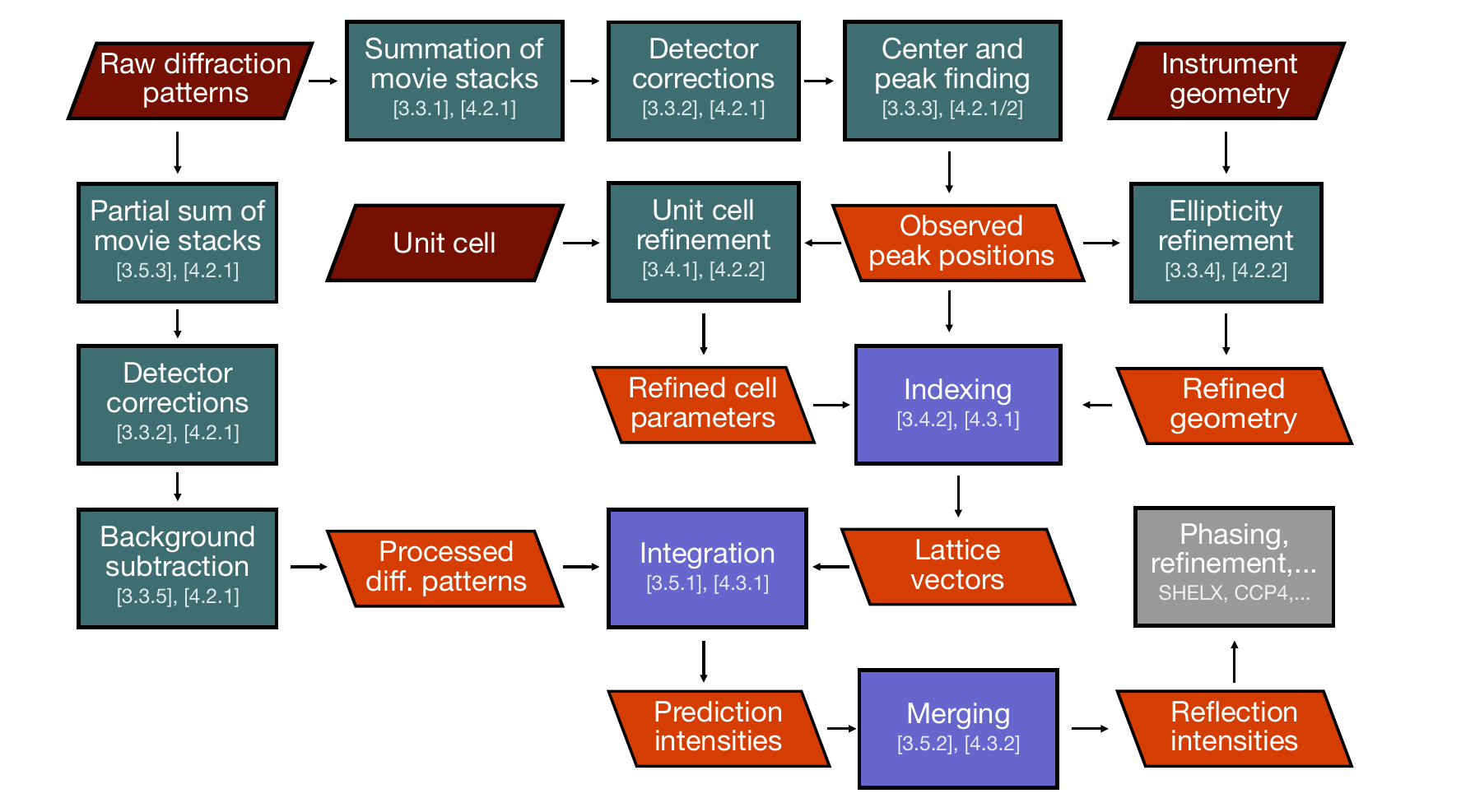}
    \caption{Journey of SerialED data through our data reduction pipeline.
    Green and blue boxes represent processing steps conducted in \emph{diffractem} and \emph{CrystFEL}, respectively; section numbers in this paper relating to each step are indicated.
    Red and orange parallelograms represent input data and important intermediate results, respectively.
    The final result (reflection intensities) is then handed over to structure solution software (grey box).}
    \label{fig:flowchart}
\end{figure*}

\begin{figure*}
    \centering
    \includegraphics[width=0.95\textwidth]{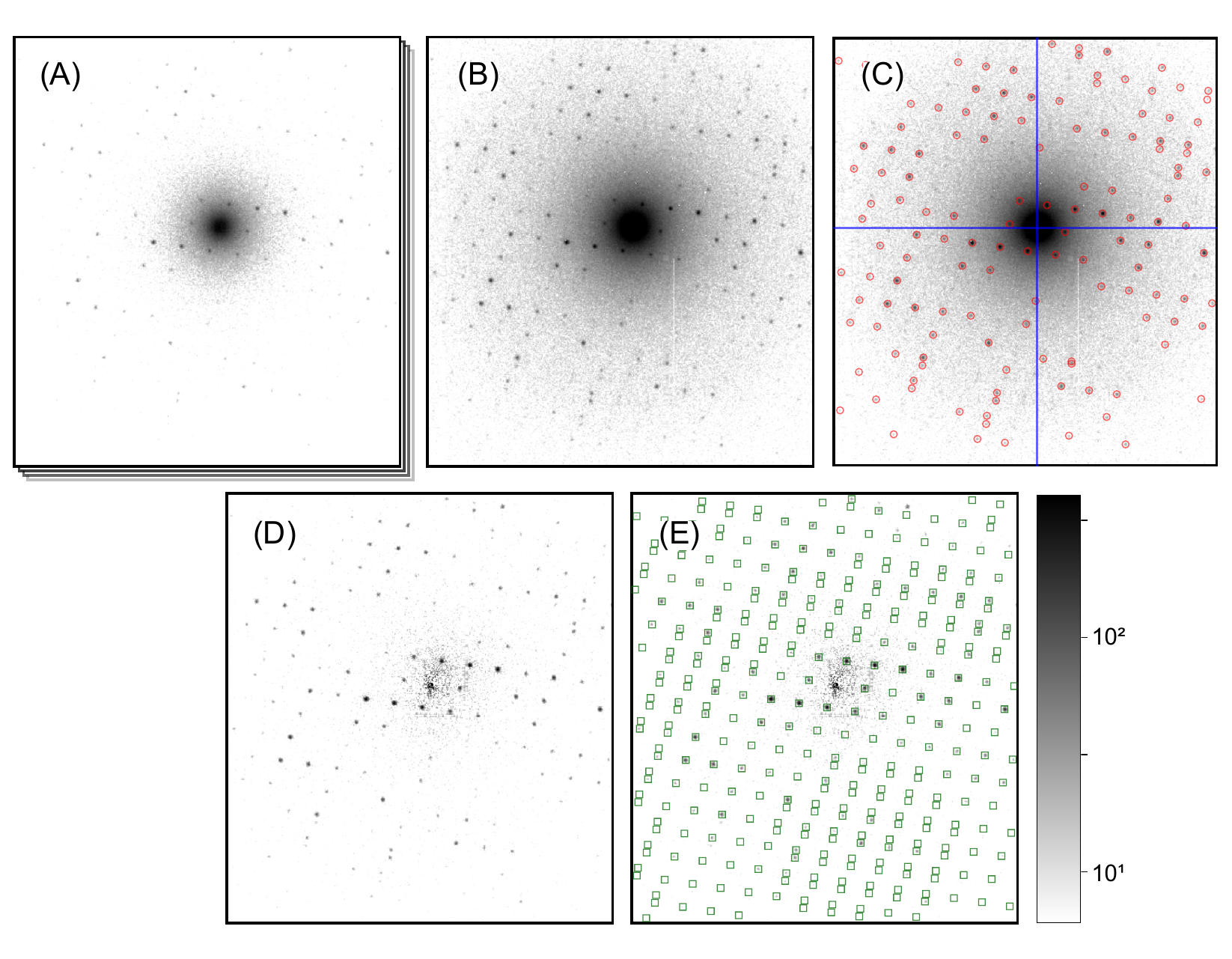}
    \caption{Processing steps of a single diffraction pattern. 
    All patterns are shown on the same, logarithmic scale. 
    (A) Initial dose-fractionation stack (first fraction is shown).
    (B) Aggregated pattern over a range of four dose-fractionation frames.
    (C) Pattern center (blue cross-hair) and Bragg peaks (red circles) have been determined.
    (D) Aggregated pattern after background subtraction.
    (E) Predicted Bragg reflections (green squares) have been computed after successful indexing.
    In the integration step, those will be included as single observations.}
    \label{fig:StepByStep}
\end{figure*}

In this section, we will describe the essential steps of a SerialED data processing pipeline, starting from a set of recorded diffraction patterns to merged reflection intensities, which can then be exported to standard software for phasing and refinement, such as \emph{PHENIX}~\cite{Adams2010}, \emph{CCP4}~\cite{Winn2011}, or \emph{SHELX}~\cite{Sheldrick2010}.
While a large portion of steps to process serial crystallography data have been addressed in established packages such as \emph{CrystFEL}~\cite{White2012}, \emph{cctbx.xfel}~\cite{Hattne2014}, and \emph{nXDS}~\cite{Kabsch2014a}, SerialED processing requires some more specific steps, which we will discuss in more detail.
As example data set from which the figures and results shown in this paper are derived, we use that taken from tetragonal hen egg-white lysozyme crystals, as has been published in~\cite{Buecker2020} (PDB-ID: 6S2N).
A flow-chart of the process is shown in Figure~\ref{fig:flowchart}; processing steps are further illustrated for a representative diffraction pattern in Figure~\ref{fig:StepByStep}. 
For the more technical details of the processing pipeline, we refer to Section~\ref{sec:diffractem}, where the practical use of our processing program package \emph{diffractem} in conjunction with the serial crystallography package \emph{CrystFEL}~\cite{White2012,White2019} is discussed, and the Jupyter notebooks supplied as supplementary material.

\subsection{Pre-processing}
\label{sec:preprocessing}

We start by applying several pre-processing steps to the diffraction patterns, that is, aggregation of dose-fractionation stacks, correction of artifacts introduced by the detector, accurate determination of the pattern center (zero-order peak) and position of Bragg peaks, and general handling of metadata.

\subsubsection{Sorting and aggregation}
\label{sec:aggregation}

The first processing step is to reject superfluous shots (i.e., single exposures on the camera), which might be present in the dataset due to auxiliary scan points inserted during data collection~\cite{Buecker2020} to mitigate hysteresis effects during beam scanning.

Next, if dose-fractionation movies have been collected where several images correspond to the same diffraction pattern from a single, still crystal (Figure~\ref{fig:StepByStep}~A), the successive frames are summed over an arbitrary number of frames adding up to an equivalent exposure time, as to provide a reasonable balance between signal-to-noise ratio of low-resolution peaks and pattern resolution (which fades at later times) for each crystal (Figure~\ref{fig:StepByStep}~B).
As most of the processing steps, such as peak finding and indexing, are independent of the exact peak intensities affected by damage effects, this choice of equivalent exposure is not critical at this point, as long as the diffraction peaks are well visible.
The optimal exposure time can be exactly determined during the later steps of peak integration and merging (Section~\ref{sec:fractionation}).

\subsubsection{Detector artifact correction}
\label{sec:detector}

Any real electron detector shows a range of imperfections, three of which we account for during processing:
\begin{itemize}
    \item Faulty pixels, which yield zero, extremely high, or excessively fluctuating values, are a primary source of errors during peak finding, indexing, and integration.
    In our processing pipeline, we assume the existence of an accurate dead-pixel map, i.e., an image file with defined pixel values at faulty or intact pixels, respectively, which can be obtained by recording images with even illumination.
    During processing of diffraction patterns, the values of these pixels are overwritten by interpolation from adjacent pixels, or (at the user's choice) flagged for exclusion from further processing steps.
    \item The response of a detector to a homogeneous illumination (flat-field) is typically non-uniform.
    If the raw data are not corrected for this effect already, this can be accounted for by a simple normalization during processing.
    \item For high pixel values, a detector can saturate, in ways which may differ between various models.
    Integrating detectors such as CCD, indirect CMOS, or linear-mode direct detectors saturate in the total counts per pixel with a sharp cut-off, which can be treated by exclusion from further analysis steps, in a similar way to dead pixels.
    On the other hand, counting detectors e.g. of counting-mode direct or hybrid-pixel type, suffer from continuously increasing coincidence-loss saturation as a function of count \emph{rate}.
    For the latter, if previously characterized, a saturation model can be applied.
\end{itemize}

Our example data set has been recorded using a hybrid-pixel detector with a large number of dead pixels, which have to be taken into account, but a fairly even flat-field.
The used count rates range into the saturation range near the center of the diffraction pattern (i.e. close to the transmitted beam), which is accounted for by using a paralyzable dead-time model~\cite{Feller2015}, parametrized from independent measurements.
All of those corrections are applied before any further image analysis.

\subsubsection{Pattern centering and peak finding}
\label{sec:centering}

\begin{figure}
    \centering
    \includegraphics[width=0.95\columnwidth]{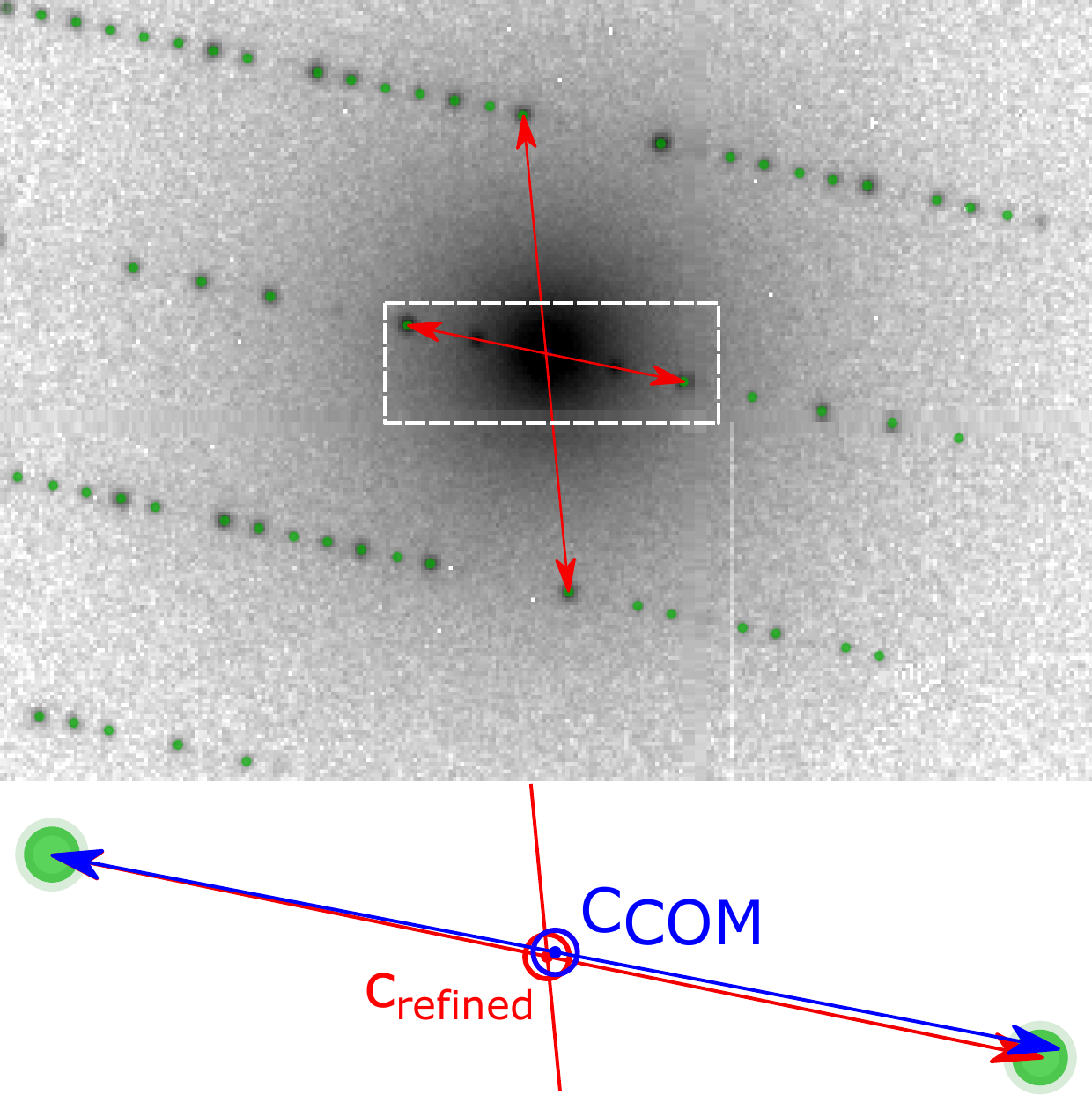}
    \caption{Pattern center refinement from Friedel mates.
    In electron diffraction patterns, even away from zone axes, a large number of Friedel mates is simultaneously visible, such as those marked by arrows.
    As they are symmetric about the zero-order beam, the positions of each pair can be used to refine the initial estimate $\vec{c}_\mathrm{COM}$ of the pattern center to the more accurate $\vec{c}_\mathrm{refined}$.
    }
    \label{fig:friedel}
\end{figure}

For successful indexing of the diffraction patterns, it is of crucial importance to accurately know the center (zero-order beam) position of each diffraction pattern.
For serially collected data, where the beam is moving between crystals, the pattern center tends to fluctuate between beam positions due to residual alignment issues of the microscope (beam-shift pivot).
Hence, the beam center has to be found for each pattern separately, which in our pipeline is tightly coupled to the detection of Bragg peaks.

To find both the pattern center and peak positions, first we determine the center of mass of the diffraction pattern, excluding all pixels whose values fall below a threshold chosen such that only a region around the center is taken into account.
Next, we apply a two-dimensional least-squares fit of a Lorentzian function to a region of tens of pixels of diameter around the found center of mass position, to obtain a more accurate estimate for the pattern center.
Peaks are now detected using the \emph{peakfinder8} algorithm~\cite{Barty2014}, which inherently takes into account a radially symmetric background as typically present in electron diffraction patterns due to multiple inelastic/elastic scattering~\cite{Latychevskaia2019}.

To further refine the center position of each electron diffraction pattern, we make use of the fact that, due to the flat Ewald sphere of electrons, even for patterns away from a zone axis, many Friedel-mate pairs (with Miller indices h,k,l and -h,-k,-l, respectively) can be found in a single image, as shown in Figure~\ref{fig:friedel}.
Each pair is necessarily symmetric with respect to the pattern center, which can be used to further refine the estimate of the pattern center.
The refinement is performed by defining a score function:
\begin{align*}
    F(\mathbf{r}_0) = \frac{1}{2N_\mathrm{pk}}\sum_{i,j}^{N_\mathrm{pk}} \exp\left[-\frac{1}{2\sigma^2}(\mathbf{r}_i + \mathbf{r}_j - 2\mathbf{r}_0)^2\right],
\end{align*}
with all found peaks at pixel position $\mathbf{r}_i=(x_i, y_i)$ of characteristic width $\sigma\approx 2$~pixels, and performing a least-squares minimization on $F^{-1}$ in order to obtain the refined pattern center at $\mathbf{r}_0=(x_0, y_0)$ with sub-pixel accuracy.
We find that further refinement as performed by pattern indexing codes does not lead to any significant improvement.
In Figure~\ref{fig:StepByStep}~C, a typical result of pattern centering and peak finding is shown.

\subsubsection{Ellipticity finding}
\label{sec:ellipticity}

A common artifact introduced by the electron optics in an electron microscope column is a slight elliptical distortion of the diffraction pattern which, even in the range of only few percent, can severely hamper the efficiency of crystallographic algorithms.
Hence, care has to be taken to account for the distorted geometry, especially during the indexing and integration steps.

\begin{figure}
    \centering
    \includegraphics[width=0.95\columnwidth]{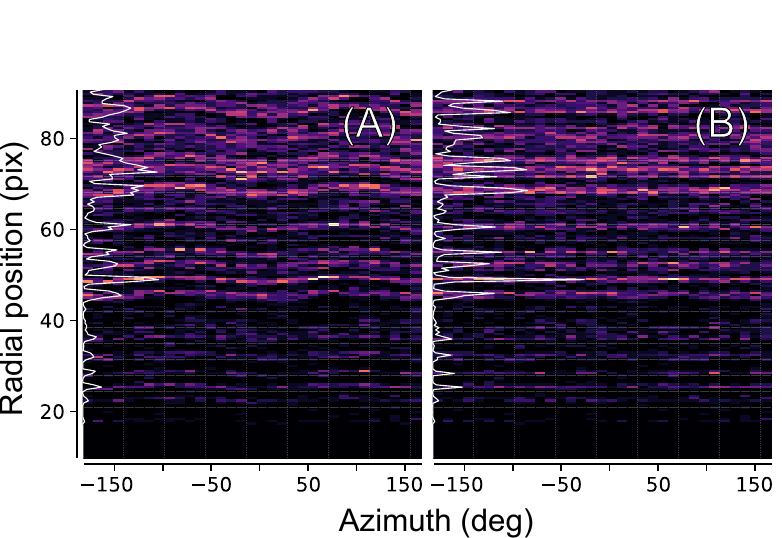}
    \caption{Ellipticity refinement. In order to correct for the elliptical distortion of diffraction patterns introduced by electron optics, we histogramize all found diffraction peaks (from all images) in two-dimensional polar coordinates.
    (A) Elliptical distortion manifests itself in a modulation of the position of major features near the inverse layer spacings of the crystal.
    Azimuthal integration into a radial profile (white line) yields a blurred, low-contrast pattern.
    (B) Same, after correcting the positions of the peaks according to an elliptical model before histogramization.}
    \label{fig:geomrefine}
\end{figure}

The ellipticity can be derived from the data itself, by computing a two-dimensional histogram of \emph{all} measured diffraction peak positions (relative to the pattern center) in radial coordinates, as shown in Figure~\ref{fig:geomrefine}.
In an ideal geometry, there is no dependence of any features (virtual powder rings) on the azimuth angle.
The elliptical distortion as seen in Figure~\ref{fig:geomrefine}~A can hence be computed by iteratively modifying the peak positions according to their azimuth angle, and recomputing, until no dependence is found anymore as seen in Figure~\ref{fig:geomrefine}~B.

\subsubsection{Background rejection.}
In contrast to X-rays, inelastically scattered electrons are not removed from the beam, but continue their trajectory toward the detector, thus appearing in the recorded data unless an energy filter is used.
While the differential cross section for inelastic scattering drops off quickly at angles small compared to typical Bragg reflection angles, \emph{combined} elastic and inelastic scattering leads to a pronounced, radially symmetric background~\cite{Latychevskaia2019} in unfiltered electron diffraction patterns.
As long as the peak integration algorithm, which serves to extract the summed intensity of each peak from the images, can handle this background appropriately, it in principle does not impact the obtained values, apart from a decreased signal-to-noise ratio at low resolutions.
However, we find that subtraction of the radially symmetric background not only aids to visually represent and assess the diffraction patterns as seen in Figure~\ref{fig:StepByStep}~D, but also simplifies the peak integration process (due to the absence of a background gradient) and leads to more consistent results after merging.
The tools provided by \emph{diffractem} as described in Section~\ref{sec:proc2d} allow to reject any radially symmetric signal following a prescription as follows:
\begin{enumerate}
\item Computation of the radial profile of the inelastic background by azimuthal averaging around the previously found pattern center, excluding a generous area around each of the found Bragg spots to avoid over-correction.
\item Median filtering of the profile to reduce noise and reject residual ripple caused by weak, unidentified Bragg peaks.
\item Computation of expected background image from profile by assigning pixel values based on the radius with respect to the pattern center.
\item Subtraction of the computed background from the actual diffraction pattern.
\end{enumerate}
We find the outcome of this procedure to be satisfactory even for the dense diffraction patterns of proteins.


\subsection{Indexing}

After corrected diffraction patterns with annotated center and peak positions have been computed (Figure~\ref{fig:StepByStep}~C), the next step is indexing the patterns, that is, deriving the unit cell parameters, which are assumed to be narrowly distributed over all crystals, and the orientation of each individual crystal.
Common processing pipelines for serial X-ray diffraction data solve the indexing problem by estimating a unit cell for each pattern separately, and if necessary, iteratively refining the obtained solutions~\cite{White2019}.
However, owing to the short de-Broglie wavelength of high-energy electrons, diffraction patterns are almost entirely devoid of three-dimensional information, which precludes the required determination of the crystal unit cell from single diffraction patterns.
While approaches exist to bootstrap the cell information from all patterns taken as a whole~\cite{Jiang2009}, the cell can also be experimentally derived from ancillary rotation-based data or multi-tilt serial data (publication in preparation).
We defer the discussion of cell-finding to future publications, and instead focus on two remaining steps of indexing, namely, accurate refinement of the unit cell parameters, and determination of the orientation of each individual crystal.

\subsubsection{Unit-cell refinement}
\label{sec:peak_refine}

\begin{figure*}
    \centering
    \includegraphics[width=0.95\textwidth]{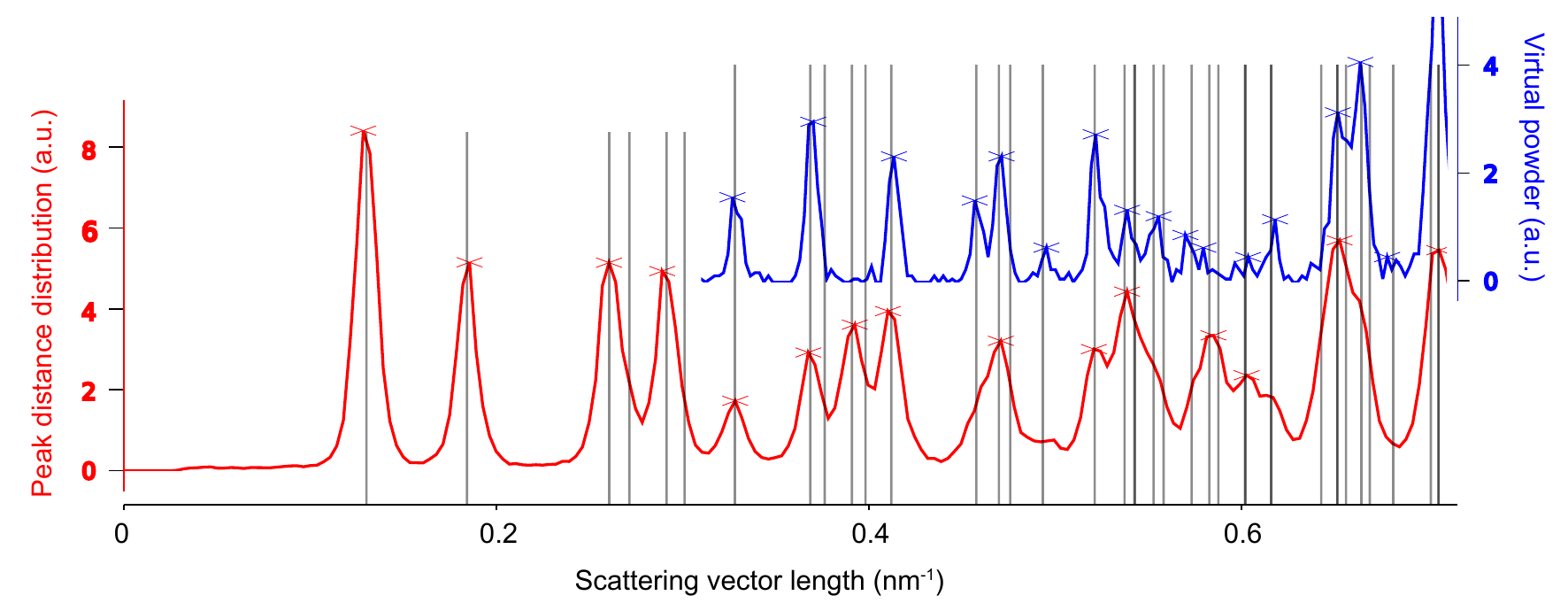}
    \caption{Unit-cell refinement of the example system's tetragonal cell with $a=\SI{78.9}{\angstrom}, c=\SI{37.9}{\angstrom}$.
    After peak finding, the scattering vector length, and the pair-wise distances between the peaks (within each pattern) are histogramized over all patterns (blue and red curves, respectively), yielding peaks at the inverse layer spacings of the crystal.
    The unit cell can accurately be refined by fitting the computed layer spacings (grey lines) to the observed peaks in both distributions.}
    \label{fig:cellrefine}
\end{figure*}

If the Bravais lattice and reasonable estimates of the cell parameters are known, the latter can be refined against radial distribution functions derived from the found peaks in the \emph{entire} data set, as shown in Figure~\ref{fig:cellrefine}.
To this end, we consider two types of peak information, both of which are histogramized with respect to their radial coordinate.
Firstly, we simply consider the radial position of peaks with respect to the pattern center, which can be related to the Bragg angle $2\theta$ and hence the crystal's inverse layer spacings.
The according histogram is known as a \emph{virtual powder pattern}, as it effectively corresponds to a super-resolution measurement of a powder diffraction pattern.
Secondly, we compute the distribution of all pair-wise distance vectors between peaks present in each pattern.
Due to the small Bragg angles of electrons (paraxial regime), the lengths of those similarly match inverse layer spacings, which can then be averaged over the entire dataset. 
The advantage of the second method is that the result displays pronounced peaks near the primitive-cell basis vectors, which are hardly or not at all (due to systematic absences) present in the virtual powder pattern.
Once the distribution functions are computed, the cell parameters are refined against them by matching the predicted layer spacings to their respective peaks.

\subsubsection{Indexing using \emph{pinkIndexer}}
\label{sec:indexing}
Now that the unit cell of the crystals is known, the orientation of each crystal with respect to the experiment geometry can be determined by an exhaustive search over all possible rotations, for which several implementations are available~\cite{Ginn2016, Beyerlein2017, Smeets2017, Li2019, Gevorkov2019}.
We use \emph{pinkIndexer}~\cite{Gevorkov2019}, which has been tested extensively on electron data, and is directly integrated into the \emph{indexamajig} program of~CrystFEL.
Before running the indexing process, it may be required to screen the parameters of the indexing on a small sub-set of diffraction patterns, which should be selected by the number of found peaks and visual appearance.
Factors impacting the successful indexing rate are the accuracy of unit cell parameters, proper centering of the patterns, sampling density of rotational space, and the assumed radius of Bragg spots in reciprocal space.
While the first two can be refined using the methods described above, the others have to be found heuristically for the sample under study.
While the sampling density depends critically on the unit cell size, the optimal setting for the Bragg spot radius is defined by the interaction region between the electron beam and continuous crystalline blocks, which can be limited by crystal size, beam size, or crystal mosaicity and bending~\cite{Gallagher-Jones2018}, as well as the beam convergence angle.
Given the typical parameters of three-dimensional electron diffraction, realistic values are below \SI{0.005}{\per\angstrom}.
While a too large value tends to assign patterns to a near-zone-axis geometry with densely packed peaks, a too small value can preclude any successful indexing.

Depending on the sampling density used for the orientation search, up to one minute of computation time is required for each crystal; however it is straightforward to distribute the calculation over arbitrarily many processor cores on a cluster system, which is automated in our processing software~\ref{sec:indexamajig}.

\subsection{Peak integration, merging, and validation}

Having determined the orientation of each crystal, we can proceed to integration and merging, that is, from the manifold of indexing solutions deriving a complete set of estimates for the Bragg spot intensities for firstly individual patterns (integration), and secondly the entire dataset (merging).

\subsubsection{Integration of intensities from indexing results}
\label{sec:integration}

The unit cell vectors of each diffraction pattern as found by the indexing are used to extract the intensities of observations of Bragg reflections, which may be partial~\cite{White2014}.
To accomplish this \emph{peak integration} step, we use functionality built into CrystFEL, as outlined in more detail in Section~\ref{sec:indexamajig}.
Briefly, the positions of all Bragg reflections that could reasonably be present in each diffraction pattern are computed (spot \emph{predictions}) from the crystal orientation and a refined reciprocal spot radius, as shown in Figure~\ref{fig:StepByStep}~E.
Then, the pixel intensities around each prediction position are integrated, using one out of several available methods such as profile fitting~\cite{Rossmann1979} and simple summation within an appropriately chosen radius~\cite{White2013}.
We usually find the simplest method, that is, summation without any additional refinement steps, to be most effective; background-gradient correction as also offered is only required, if the diffraction patterns are not background-subtracted.

\subsubsection{Merging and validation of integrated intensities}
\label{sec:merging}

\begin{figure*}
    \centering
    \includegraphics[width=0.95\textwidth]{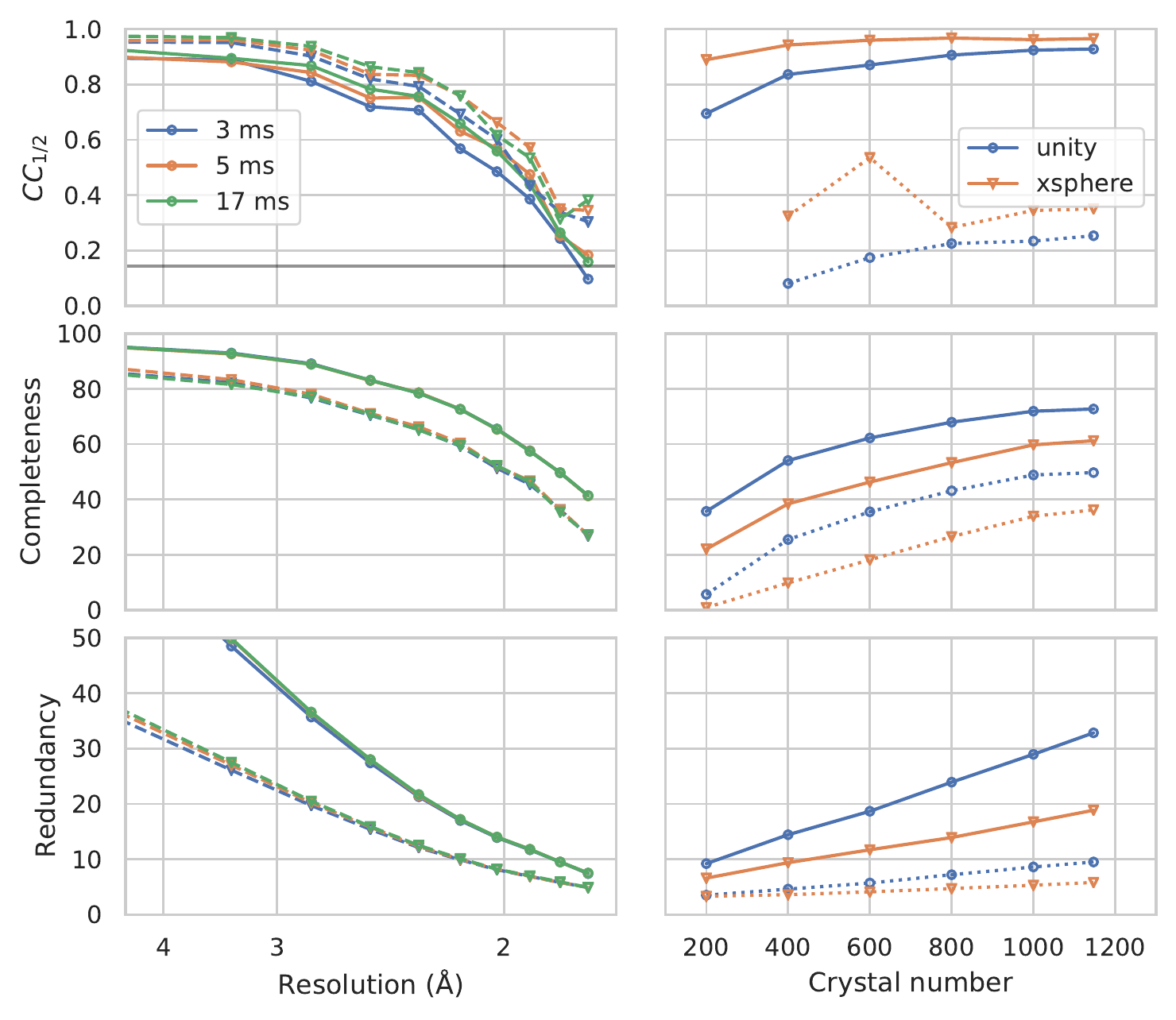}
    \caption{
    Merging statistics as a function of resolution (A-C) and crystal number (D-F): Half-set Pearson correlation $CC_{1/2}$, dataset completeness, and observation redundancy.
    In (A-C), results are shown for three different integration times in different colors, and for merging without (solid lines, circles) or with (dashed lines, triangles) the \emph{xsphere} partiality model~\cite{White2014}.
    In (D-F), solid and dashed lines represent overall values (entire resolution range) and those from the second-highest resolution shell as shown in (A-C), which is centered at \SI{1.85}{\angstrom}.
    Blue circles and orange triangles represent results from merging without and with partiality modeling, respectively; iterative scaling was enabled in both.}
    \label{fig:merging}
\end{figure*}

After the measured Bragg spot intensities from all shots are extracted and stored, they have to be merged into a full crystallographic data set.
Firstly, depending on wether the crystal's space group shows an indexing ambiguity, it needs to be resolved using, which can be performed using a clustering algorithm~\cite{Brehm2014,Kabsch2014a,White2016}, which is provided as a part of CrystFEL (\emph{ambigator}).
Then, the many observations of each Bragg reflection are combined, in the simplest case by averaging without further weighting (``Monte-Carlo method'').
Additionally, iterative global and resolution-dependent scaling can be introduced, which leads to significant improvements~\cite{White2016}.
Finally more elaborate models for merging, which explicitly model the amount partiality of each observation, are available, which in our experience leads to varying, sample-dependent results (Figure~\ref{fig:merging}); a detailed discussion of partiality modeling for SerialED will be the subject of future work.

In order to assess the overall statistics and quality of the merging result (and hence, the effectively the entire data reduction pipeline), it is of crucial importance to evaluate some important validation metrics.
While traditional merging quality indicators such as $R_\mathrm{merge}$ are inadequate for serial dataset due to their strong partiality~\cite{White2012}, the half-set Pearson correlation coefficient between Bragg intensities merged form a half-sets of the crystals $CC_{1/2}$~\cite{Karplus2012} provides a robust figure of merit for consistency of the dataset.
Furthermore, the completeness of the dataset, as well as the mean number of observations of each reflection (redundancy) are of highest concern.
In Figure~\ref{fig:merging}, these quantities are shown for our example data set, as a function of resolution shell, and of the number of merged crystals (by picking a sub-set from the data).
We can observe that the correlation coefficient, which is near-unity at low resolution, drops below a threshold of 0.143 (corresponding to $CC^*=0.5$~\cite{Karplus2012}) at about 1/(\SI{1.8}{\angstrom}), which is hence a reasonable resolution cut-off for phasing and refinement steps.
Another important observation is that the completeness of the dataset appears to converge to a value significantly less than 100\% when increase the crystal number; this clearly indicates the presence of preferred crystal orientation, which cannot be significantly mitigated by increasing the number of crystals.
Such preferred orientation issues can however be mitigated using a tilt of the sample stage or specifically prepared sample grids~\cite{Wennmacher2019}.

\subsubsection{Processing of dose-fractionation movies.}
\label{sec:fractionation}

If a sufficiently fast diffraction detector is available, it is advisable to collect SerialED in dose-fractionation mode, that is, taking a series of frames (movie) for each crystal in rapid succession, as shown in Figure~\ref{fig:StepByStep}.
This technique is commonly applied in single-particle microscopy, and while motion blur is not of concern for diffraction data, dose fractionation allows to select an optimal exposure time, and hence radiation dose per crystal \emph{a posteriori}~\cite{Buecker2020}.
Assuming that the orientation of crystals does not significantly change between the movie frames, and hence the indexing solution is valid for all frames equivalently, the exact choice of considered integration time is mostly irrelevant up to the point of integration, as long as the visible Bragg peaks at low to intermediate resolution can be reliably found.
It is only in the final steps that results should be derived for different integration time separately.
This can be accomplished by ``broadcasting'' the position of stop predictions to a dataset that comprises diffraction patterns with varying aggregation length as described in Section~\ref{sec:aggregation}, and re-running integration and merging on those sets.
Our analysis programs provide convenient functions to automate this process and guide the user to an optimal choice of exposure time.

\section{Implementation of SerialEM Processing}
\label{sec:diffractem}

The various steps of data processing explained in the previous sections can be performed in our Python software package \emph{diffractem}, which provides the necessary functionality directly, or via tight integration with \emph{CrystFEL} through wrapper functions.
Besides a few command-line tools, diffractem is intended for comfortable use within Jupyter notebooks, a common platform for scientific data analysis and data science in general.
This section will introduce some key concepts of diffractem.
For more in-depth examples and explanations, we refer the reader to the annotated Jupyter notebooks provided as supplementary information to this paper.

\subsection{Data structures and file format}
\label{sec:dataset}

Diffraction images and meta data are accessed and managed via instances of diffractem's \inlinecode{Dataset} class.
A single \inlinecode{Dataset} object represents arbitrarily many data files that each correspond to a SerialED acquisition run from one grid region.

\subsubsection{Data hanlded by \emph{diffractem}}
In diffractem's terminology, a \emph{shot} corresponds to a diffraction pattern recorded by the detector (equivalent to an \emph{event} in CrystFEL), whether it constitutes a hit on a crystal or not.
If dose-fractionation is used, the many shots obtained from the same crystal are referred to as the \emph{frames} of said crystal.
In SerialED, thousands of raw diffraction patterns can be acquired per hour. 
Thus, the initial raw data comprises a large number of 2D diffraction patterns (shots), forming together a 3D data cube referred to as \emph{image stack} in the following, with associated meta-data.
Such meta-data can be defined per diffraction pattern (\emph{shot table}), per crystal (\emph{feature table}), or per grid region (\emph{global meta-data}).
The number of peaks in a diffraction pattern, the position of a crystal on the sample grid, and the camera length setting are examples of per-shot, per-feature, and global data, respectively.
The metadata can extensively be changed and extended along the data processing pipeline, where the \inlinecode{Dataset} object ensures consistency of image and meta data. 
Destructive processing steps that either change actual image data (such as background correction) or remove shots are handled by generating a new, modified \inlinecode{Dataset} object.

Within the \inlinecode{Dataset} object, the shot and feature tables are accessible as \emph{pandas} DataFrame objects~\cite{Pandas} via the attributes \inlinecode{Dataset.shots} and \inlinecode{Dataset.features}.
Their number of rows always correspond to the number of shots and crystals stored in the \inlinecode{Dataset} object, respectively.
On the other hand, the number of columns is arbitrary, and commonly increases once new per-pattern analysis results become available.
In any case, key columns such as the file name and location of diffraction data in the image stack, as well as the sample name, and identification numbers of each crystal and grid regions have to be present.
Global meta-data (typically comprising instrument parameters such as camera length or exposure time) can be accessed or directly merged into the shot table using the \inlinecode{Dataset.merge\_meta} method.

\subsubsection{Stacks and memory management}
\label{sec:stacks}

The image stack comprising the actual diffraction data is often too large to fit into the main memory of a typical mid-range workstation computer.
Hence, to manage this amount of data and the ensuing parallel computations, we employ the \emph{dask} package~\cite{Dask}, which allows to transparently access larger-than-memory data arrays from disk, and to build lazy computation pipelines, that can be executed efficiently in parallel (see supplementary Jupyter Notebooks for details).
A \inlinecode{Dataset} object can contain an arbitrary number of N-dimensional dask arrays (which behave analogously to \emph{NumPy} arrays), referred to as \emph{stacks}, the first dimension (dimension 0) of which must always equal the number of shots contained in the \inlinecode{Dataset} (and hence the number of rows in the shots table).
Besides the actual diffraction data stack (constituting a three-dimensional stack), typical stacks in a \inlinecode{Dataset} object are the data of found diffraction peaks in each image, in \emph{CXI} format~\cite{Maia2012}.
Generally, data stacks can be added or overwritten using the \inlinecode{Dataset.add\_stack} method and accessed via attributes of the form \inlinecode{Dataset.<stackname>}.

\subsubsection{Slicing, selecting, and aggregating data}

A common task during the preprocessing of a diffraction data set is to reject shots based on criteria such as a minimum number of Bragg peaks or a maximum level of background signal.
Such selections can easily be performed using the \inlinecode{Dataset.get\_selection} method, which allows for selections of sub-sets via query strings acting on columns of the shot list.
As an example, the code line \inlinecode{ds\_sel = ds.get\_selection('num\_peaks >= 15')} generates a new \inlinecode{Dataset} object \inlinecode{ds\_sel}, containing only shots from \inlinecode{ds} where more than 15 Bragg peaks have been detected.
In this step (as in all other methods of \inlinecode{Dataset}), it is ensured that all stacks and tables are kept consistent.
The related method \inlinecode{Dataset.aggregate}, which accepts a similar query string, will, on top of slicing, apply different group-wise aggregation functions to the data stacks, or a subset thereof; its typical application of the summation of dose fractions of the same diffraction pattern as described in Section~\ref{sec:fractionation}.
Please see the supplementary Jupyter notebooks for more detailed explanations and examples.

\subsubsection{Diffractem data files}
\label{sec:file_format}

\begin{figure*}
    \centering
    \includegraphics[width=0.95\textwidth]{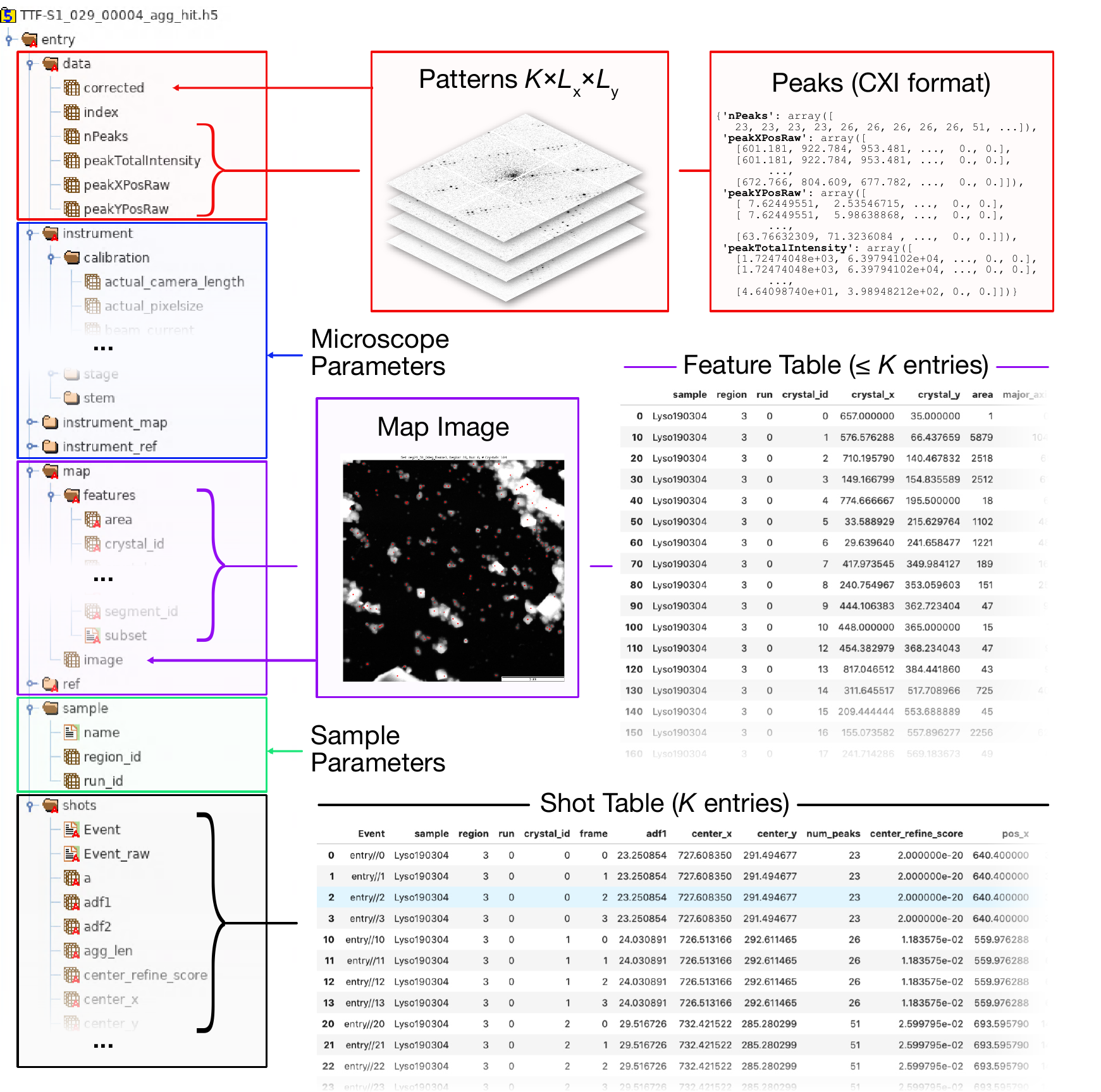}
    \caption{Structure of HDF5 file as used by \emph{diffractem}.
    Typically, each HDF5 holds data from a single SerialED run on one grid region; a \inlinecode{Dataset} object typically manages data from many such files, automatically concatenating all information.
    On the left side, a tree view of the internal folder/dataset hierarchy of a HDF5 file is shown.
    On the right side, various types of information and (via arrows and braces) their location within the HDF5 file are shown.
    In compliance with the \emph{NeXus} convention, all data is stored under a global \inlinecode{/entry} group.
    Explanations of the top-level groups (A) \inlinecode{data}, (B) \inlinecode{instrument}, (C) \inlinecode{map}, (D) \inlinecode{sample}, and (E) \inlinecode{shots} are given in the main text.
    }
    \label{fig:file_format}
\end{figure*}

Diffractem stores its data in \emph{HDF5} files largely follwing the \inlinecode{NeXus} convention~\cite{Konnecke2015}, which is becoming a common standard in X-Ray crystallography, and can by now be processed by most crystallography libraries.
The data within the files can be accessed from all common programming languages through bindings of the HDF5 library, such as \inlinecode{h5py} for Python, and can directly be mapped into larger-than-memory arrays using the dask package, as described above.
Each file holds data from a continuous acquisition run on a single region on the sample, corresponding to a single map image as shown in Figure~\ref{fig:serialed}~A on which crystals have been identified prior to diffraction data collection.
A multitude of acquisition runs from the same sample which shall be analyzed as a whole can be defined using simple text files with corresponding HDF5 file names on each line, and a \inlinecode{.lst} extension by convention.
Using the \inlinecode{Dataset.from\_files} method, data can be loaded from a single file, a list file, or a range of files implicitly defined using wildcard characters.
Both the HDF5 and list file specifications are consistent with CrystFEL.

HDF5 files are internally organized into \emph{groups} and \emph{datasets}, roughly corresponding to folders and files in a file system. 
Datasets can be arrays of arbitrary dimension, and have a uniquely assigned data type.
Mirroring the structure of a \inlinecode{Dataset} object, a diffractem data file contains primarily three types of entities:
\begin{itemize}
    \item Tabular data such as the shot list and the feature list, are stored as groups comprising one-dimensional HDF5 datasets, each corresponding to a single table column (Figure~\ref{fig:file_format}~E~and~C, respectively).
    Those tables are loaded into memory as pandas DataFrames on loading the dataset, as described in Section~\ref{sec:dataset}.
    \item Data stacks, that is, arrays with an arbitrary number of dimension, where the first (dimension 0 in Python convention) dimension corresponds to a given shot (Section~\ref{sec:dataset}), are stored as HDF5 datasets within the group \inlinecode{data} (Figure~\ref{fig:file_format}~A).
    All stacks are mapped into dask arrays when loading the dataset.
    \item Ancillary per-file instrument metadata, which can be accessed using \inlinecode{Dataset.merge\_meta} is stored in a hierarchical structures (Figure~\ref{fig:file_format}~B~and~D).
\end{itemize}

In Figure~\ref{fig:file_format}, a typical HDF5 file structure, and how it maps to the attributes of a \inlinecode{Dataset} object, is illustrated.

\subsection{Processing functions}
\label{sec:proc_functions}

In this section we describe functions that act on data stored within \inlinecode{Dataset} objects, specifically image stacks and Bragg peak data.

A commonly used ancillary tool for the functionality described in this and the next section is the \inlinecode{PreProcOpts} class contained in the \inlinecode{diffractem.pre\_proc\_opts} module. 
The attributes of this class hold values of a large number of options pertaining to the entire data processing workflow, such as which steps of the pipeline should be applied by default, but also experiment parameters such as the accurate camera length and distortion.
The attribute values of a \inlinecode{PreProcOpts} object are stored to and read from in a human-readable \inlinecode{.yaml} file, which can be continuously adjusted while working interactively on processing a dataset, and will in its final state document the exact parameters used, ensuring full reproducibility.

\subsubsection{Stack processing}
\label{sec:proc2d}
Diffractem's functions for processing image stacks as required for pre-processing (see Section~\ref{sec:preprocessing}) are contained in the \inlinecode{diffractem.proc2d} module.
Examples for such functions are \inlinecode{correct\_dead\_pixels}, \inlinecode{lorentz\_fit}, or \inlinecode{get\_peaks}.
All those take an image stack as described above (as NumPy arrays) as their first argument (with more arguments for individual options). 
They return either a processed version of the input stack (e.g. dead-pixel correction, background subtraction), per-shot data which can directly be merged into a \inlinecode{Dataset} shot list (e.g. pattern center finding, virtual detector signals), or more complex per-shot data which can be stored into stacks of a \inlinecode{Dataset} object (e.g. peak finding, azimuthal averaging).

Two special, particularly relevant functions contained in \inlinecode{proc2d} are \inlinecode{get\_pattern\_info} and \inlinecode{correct\_image}, both of which represent multi-step pipelines for getting information (such as pattern center and Bragg peaks) from each shot, and for computing processed images (having undergone e.g. dead-pixel correction and background subtraction), respectively.
In contrast to the other functions, these two act on larger-than-memory image stacks stored as dask arrays (like in a \inlinecode{Dataset} object, see Section~\ref{sec:stacks}), and have their parameters defined via \inlinecode{PreProcOpts} objects.
These two functions encapsulate computationally heavy, but independent (per-shot) steps of pre-processing, and hence are preferably using parallel execution.
This is implemented using the \emph{dask.distributed} scheduler, which besides its ease of use provides convenient real-time progress reporting via a web interface.
Please consult the supplementary Jupyter notebook \inlinecode{preprocessing.ipynb} for an example pre-processing workflow.

\subsubsection{Peak processing}
\label{sec:proc_peaks}
Another set of processing functions, acting on Bragg peak positions, is contained in the \inlinecode{diffractem.proc\_peaks} module.
This comprises functions for refinement of the zero-order peak positions (pattern center) via matching of Friedel mates (see Section~\ref{sec:centering}), getting pair-wise distances from all observed peaks (pattern autocorrelation function), and the \inlinecode{Cell} class, which provides functionality for unit-cell refinement as described in Section~\ref{sec:peak_refine}.
An example for the peak refinement workflow using a \inlinecode{Cell} object and pattern autocorrelation functions is provided in the supplementary Jupyter notebook \inlinecode{peak\_processing.ipynb}

\subsection{Integration with \emph{CrystFEL}}
\label{sec:crystfel}
For all tasks that are less specific to SerialED, but pertain to (serial) crystallography in general, diffractem provides interfaces to the CrystFEL package, in particular its central command-line tools \inlinecode{indexamajig} and \inlinecode{partialator}, as well as the validation programs for merged diffraction intensities \inlinecode{compare\_hkl} and \inlinecode{check\_hkl}.
Also, functionality to parse and manipulate \inlinecode{.stream}-files, CrystFEL's output format for pattern indexing and integration results is included.
Depend on the task at hand, diffractem either calls the executables directly, or generates the required input files and a shell script containing the corresponding function calls.
The functionality for integration with CrystFEL are mostly contained in the \inlinecode{diffractem.tools} module.
While in the supplementary Jupyter notebooks, the usage of the pertinent tools is explained in detail, here we only give a brief overview of the most important functionality, especially where deviating from the standard CrystFEL workflow.

\subsubsection{Indexing and integration}
\label{sec:indexamajig}
Indexing and integration (Sections~\ref{sec:indexing} and~\ref{sec:integration}, respecitvely) in CrystFEL are performed using the \inlinecode{indexamajig} program
As input, it requires a list of HDF5 data files (\inlinecode{.lst}) containing diffraction patterns and (optionally) peak positions, a geometry file (\inlinecode{.geom}), and a unit cell specification (\inlinecode{.cell} or \inlinecode{.pdb}).
Using the \inlinecode{tools.make\_geometry} function, the geometry file can be automatically generated from a \inlinecode{PreProcOpts} object (or, respectively, the corresponding \inlinecode{.yaml} file), which automatically handles elliptical distortion found as described in Section~\ref{sec:ellipticity}.
Similarly, the specification of a unit cell after refinement as described in Section~\ref{sec:peak_refine} can automatically be generated using the \inlinecode{export} method of a \inlinecode{proc\_peaks.Cell} object.
The \inlinecode{indexamajig} executable can be called including all pertinent options (as defined in a diffractem \inlinecode{PreProcOpts} object) using the \inlinecode{tools.call\_indexamajig} and \inlinecode{tools.call\_indexamajig\_slurm} functions, where the latter sets up intermediate files and a shell script for execution through a \emph{SLURM} queue submission system.
Optionally, those, along with the geometry, cell, and virtual data files (see below), can be packed into a \inlinecode{.tar.gz} archive for convenient uploading to a computing cluster.

Diffraction pattern indexing as described in Section~\ref{sec:indexing} requires the positions of found peaks in each diffraction pattern as its primary raw-data input.
CrystFEL's \inlinecode{indexamajig} tightly couples indexing and integration of peak intensities from image data into a single, inseparable step, as described in~\cite{White2019}.
While the file format described in Section~\ref{sec:file_format} is compatible with CrystFEL and could directly used for indexing and integration in a single run, for SerialED this approach is hampered by two prohibitive shortcomings.
First, the residual movement of the zero-order beam inherent to SerialED, even if known, cannot be natively accounted for by CrystFEL, precluding proper indexing of SerialED patterns from the Bragg reflections either found in the patterns or already stored in the files during preprocessing.
Second, SerialED requires a computationally intensive grid search approach to indexing.
Coupling indexing and peak integration into a single step hence makes it impractical to optimize the (relatively fast) integration, and would require transfer of the full dataset (as needed for integration) if indexing is offloaded to off-site computing clusters.

As shown in Figure~\ref{fig:flowchart}, diffractem circumvents these issues by not running indexing on the actual data files, but on a (single) \emph{virtual} file, which is generated using the \inlinecode{Dataset.write\_virtual\_file} method and does not carry actual diffraction data.
The virtual file, while being a fully valid diffractem and CrystFEL HDF5 file, only contains the shot list and found Bragg peaks in CXI format, which are shifted for each pattern such that position of the zero-order beam remains at the center of the detector.
\inlinecode{indexamajig} can now be run on the virtual file, yielding the indexing results (that is, the reciprocal-space lattice vectors in the laboratory frame, for each crystal found in the diffraction patterns) in \inlinecode{.stream} format.
All book-keeping to associate patterns in the virtual and actual files is transparently performed using items in the shot tables, and the \inlinecode{--copy-hdf5-field} option of \inlinecode{indexamajig}.

For peak integration, we modified CrystFEL by introducing a new option to, instead of finding indexing solutions from Bragg reflections, read reciprocal-space lattice vectors and beam shift coordinates from a plain-text \emph{solution} file (extension \inlinecode{.sol}), and proceed with the standard prediction and integration pipeline from there.
To generate the solution file from the computed indexing parameters (in \inlinecode{.stream} format), the method \inlinecode{Dataset.get\_indexing\_solution} can be used, which transparently handles the case of integrating patterns that have been computed from a different range of movie frames (see Section~\ref{sec:aggregation}) than that initially used for indexing.
For the more simple case where the data that shall be integrated is identical to those that were used to generate the indexing solution, the command-line tool \inlinecode{stream2sol} which is included in diffractem, can be used alternatively.

Please see the supplementary notebook \inlinecode{indexing.ipynb} for a detailed step-by-step guide of indexing and integration.

\subsubsection{Merging and validation}
\label{sec:partialator}

The merging of single Bragg peak observations from all recorded diffraction patterns as described in Section~\ref{sec:merging} is performed using the \inlinecode{partialator} command-line program contained in CrystFEL.
Diffractem includes a corresponding wrapper function \inlinecode{tools.call\_partialator}.
It provides a convenient way to generate \inlinecode{partialator} calls from within Jupyter notebooks, also providing options to run different merging settings (e.g., with and without post-refinement or resolution cut-offs) in parallel or sequentially, optionally generating a script for submission to a \emph{SLURM} cluster queue submission system.

Finally, the merged intensities contained in \inlinecode{.hkl} files can be analyzed from Jupyter notebooks by wrapping CrystFEL's \inlinecode{check\_hkl} and \inlinecode{compare\_hkl} command-line tools into the \inlinecode{tools.analyze\_hkl} function, which provides means to automatically validate the results of many different integration and merging parameters in parallel, and wraps the results in \emph{pandas} DataFrames.
Please see the supplementary notebook \inlinecode{merging.ipynb} for an example of the merging and validation steps.

\subsection{Displaying data}
\label{sec:viewers}

\begin{figure*}
    \centering
    \includegraphics[width=0.95\textwidth]{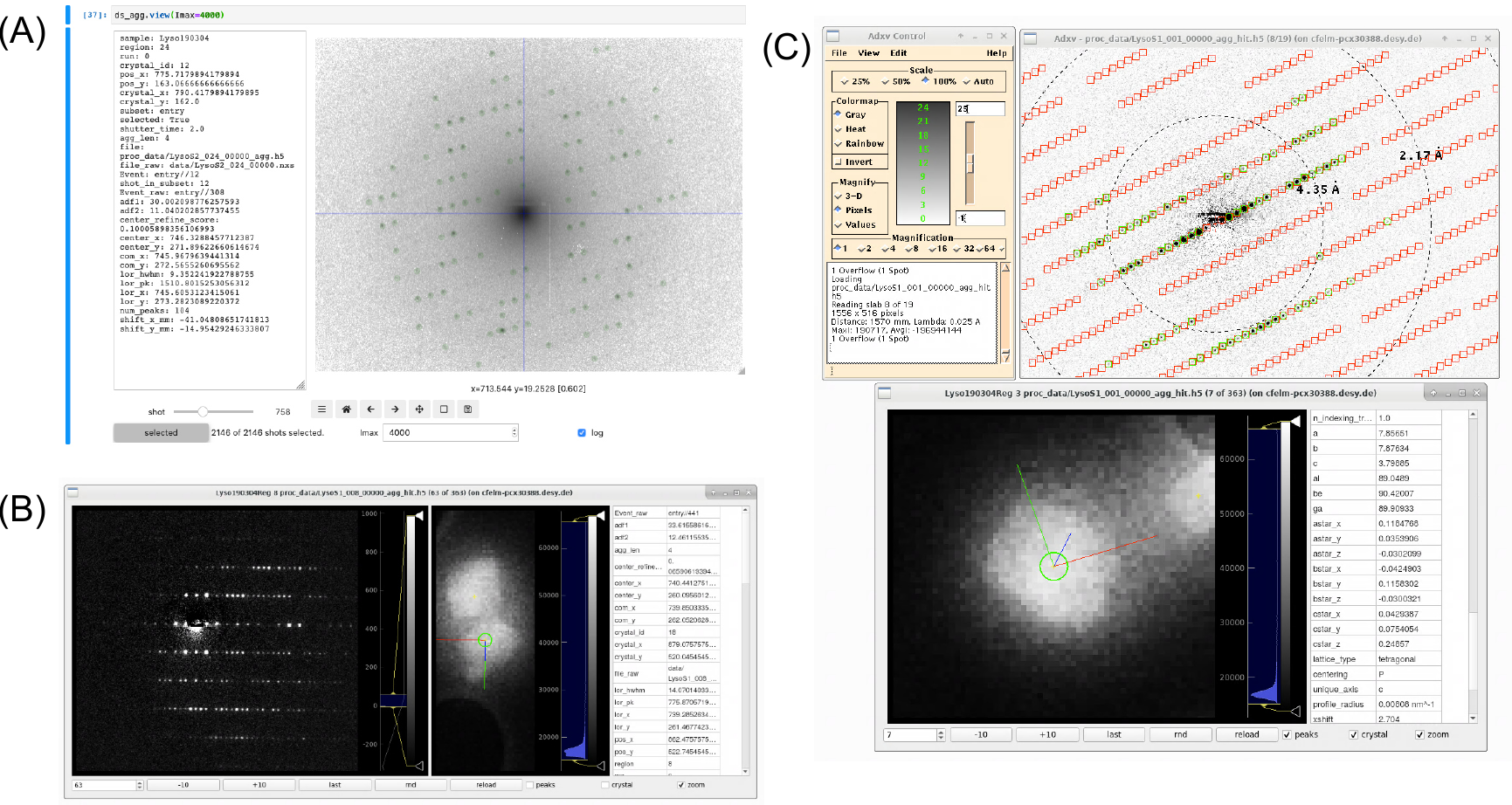}
    \caption{Screenshots of diffraction viewing tools of \emph{diffractem}.
    (A) \inlinecode{Dataset.view} running as an interactive widget inside a Jupyter notebook in a web browser.
    The diffraction pattern is shown on logarithmic scale, which is particularly useful to assess the quality of pattern center and peak finding at low resolutions; the pattern center and Bragg peaks are shown as blue cross-hair and green circles, respectively.
    On the left, data from the shot table for the shown pattern are displayed.
    The controls at the bottom allow to move between shots and set display parameters.
    (B) \inlinecode{edview} running in internal-viewer mode.
    In three columns, the diffraction pattern, the map image (optionally zoomed into the shown crystal) and meta-data from the shot table \emph{and} the indexing result from a \inlinecode{.stream} file for the shown diffraction pattern are shown, respectively.
    Image controls are at the bottom.
    (C) \inlinecode{edview} in external-viewer mode, in which the diffraction pattern is displayed through \emph{adxv}~\cite{Adxv}.
    In the pattern, found peaks (green circles) and predicted Bragg spot positions from the indexing solution (red squares) are shown.
    In the bottom \inlinecode{edview} window, the corresponding crystal is shown.
    The directions of the real-space lattice vectors $\vec{a}, \vec{b}, \vec{c}$ are shown in red, green, and blue, respectively.
    }
    \label{fig:viewers}
\end{figure*}

In order to visualize datasets being processed by diffractem, two tools with markedly different scope are provided, as shown in Figure~\ref{fig:viewers}.
Firstly, the \inlinecode{view} method of a \inlinecode{Dataset} (Section~\ref{sec:dataset}) allows for quick interactive inspection of diffraction data within a Jupyter notebook, which is especially helpful for tuning of processing parameters.
Secondly, the stand-alone program \inlinecode{edview} provides a simple graphical interface to browse through SerialED data, including correlative display of mapping and diffraction data.
In Figure~\ref{fig:viewers}, screen-shots of both tools are shown.

\subsubsection{Dataset.view}
\label{sec:widget}
An interactive viewer for diffraction data can be directly used within Jupyter notebooks in the web browser, where data are being processed.
The viewer is called by invoking \inlinecode{ds.view(<\ldots>)}, where \inlinecode{ds} is a \inlinecode{Dataset} object and \inlinecode{<\ldots>} represents additional calling arguments.
The viewer shows the data stack accessible via the \inlinecode{Dataset.diff\_data} attribute (which points to the data stack containing diffraction data), and, if present as CXI-formatted data stacks, detected Bragg peaks.
Finally, if columns \inlinecode{center\_x} and \inlinecode{center\_y} are present in the shot table, the position of the pattern center (zero-order beam) is shown as a cross-hair.

Importantly, \inlinecode{Dataset.view} acts on diffraction data stored as \emph{dask} array~\cite{Dask}, which are typically not in memory, but either on disk, or not even computed yet (lazy evaluation), if the \inlinecode{Dataset} object has not been written to disk.
They are then loaded and/or computed on-the-fly for each displayed image.
This makes \inlinecode{Dataset.view} especially suitable for interactive tuning of pre-processing parameters (such as peak-finding sensitivity thresholds) on a few selected shots, before the full computation is performed.

In the supplementary Jupyter notebook \inlinecode{preprocessing.ipynb}, the use of \inlinecode{Dataset.view} is illustrated at various points.

\subsubsection{edview}
\label{sec:edview} 
The second option for displaying diffraction data is the stand-alone viewer \inlinecode{edview}, which is available from the command line after installation of diffractem.
As input to \inlinecode{edview}, single HDF5 data files, list files, multiple data files (via file wildcards), or a \inlinecode{.stream} file can be provided.
In the latter case, indexing solutions (Bragg spot predictions and real-space lattice vectors) can be displayed.
\inlinecode{edview} shows both diffraction data and, if present, the overview maps taken in the course of a SerialED data acquisition from a grid region, including an indicator to show which crystal on the map an individual pattern belongs to.
For displaying the diffraction data, either a built-in display window (via the command-line option \inlinecode{--internal}) or \emph{adxv}~\cite{Adxv}, which is controlled by \inlinecode{edview} via a local communication socket, can be used.
If indexing information is present for a given shot, the projected directions of the real-space lattice vectors $a,b,c$ (with fixed length) are overlaid on the currently displayed crystal (if ``zoom'' is checked).

\subsection{Simple on-line pre-processing using \emph{quick\_proc}}
\label{quick_proc}
While diffractem has been designed with the usage from Jupyter notebooks in mind, there may be situations where it is preferable to run the pre-processing pipeline, up to the point of aggregated, corrected, and background-subtracted images, from the command-line.
Hence, the command-line tool \inlinecode{quick\_proc} is provided by diffractem, which executes those steps according to settings defined in a \inlinecode{.yaml} file, just as for the standard processing in notebooks (Section~\ref{sec:proc_functions}).
Furthermore, \inlinecode{quick\_proc} can run in an on-line analysis mode (using the flag \inlinecode{--wait-for-files}), where it waits for new data files from the experiment to arrive, then executes the processing, and adds the newly processed files to a \inlinecode{.lst} file for use with CrystFEL or viewing using \inlinecode{edview}.
Running \inlinecode{quick\_proc -h} provides a full reference of options.

\section{Discussion and Outlook}
\label{sec:discussion}

Using the pipeline comprising \emph{CrystFEL} and \emph{diffractem} as described in this article, processing SerialED datasets of high quality becomes a straightforward exercise, and tackling more challenging cases becomes viable.
Still, there is plenty of room for future work.
Besides usability improvements for non-expert users, such as a graphical program interface for basic operations or functions for reasonable automatic adjustment of parameters for a given sample, there are more fundamental aspects which can profit from further development.
A rather obvious starting point for future work could be inclusion of a cell-finding algorithm similar to that presented in~\cite{Jiang2009}, or even an entirely new method for indexing that would be based on considering peak data from the entire dataset instead of acting on individual patterns, similarly to single-particle analysis~\cite{Scheres2012} or expand-maximize-compress algorithms in diffractive imaging~\cite{Loh2009}.
Similarly, a more systematic study of partiality modeling for electrons is required, where partiality is especially prevalent due to the small crystal sizes (and concomitantly wide rocking curves) combined with a very monochromatic beam.
Another field of study are the effects of dynamical diffraction arising from multiple scattering, which depend on subtle details that are often challenging to grasp, in particular for biological samples made from light elements~\cite{Subramanian2015,Latychevskaia2019,Nannenga2019,Gallagher-Jones2018}.
While often considered deleterious for structure solution, careful inclusion of dynamical diffraction can lead to unique insight into molecular configurations~\cite{Palatinus2017,Brazda2019} and even might be able to solve the phasing problem for electron crystallography~\cite{Donatelli2020}.
Especially regarding the latter point, SerialED can provide the unique advantage of being able to selectively solve structures from sub-sets of data containing crystals from a given size bracket only.

While there is a large scope for future developments, already in its current state of development SerialED can provide high-resolution structures of even the most demanding nano-crystalline samples~\cite{Buecker2020}.
Data analysis, while not yet as established as for rotation techniques, is becoming a more and more routine task, helped by packages such as those described in this work.
Meanwhile, the \emph{diffractem} package (as well as \emph{CrystFEL}, which provides much of the fundamental functionality) is under constant development, as to keep making SerialED data processing more efficient, powerful, and user friendly; we hence suggest to regularly check the webpage at https://github.com/robertbuecker/diffractem for updates.

\section*{Conflict of Interest Statement}

The authors declare that the research was conducted in the absence of any commercial or financial relationships that could be construed as a potential conflict of interest.

\section*{Author Contributions}

R.B. and R.J.D.M. conceived the serial electron diffraction concept.
R.B. and P.H. developed the SerialED processing pipeline.
R.B. wrote the \emph{diffractem} software.
P.H. wrote the extensions to \emph{CrystFEL} to adapt to our analysis pipeline.
R.B. and P.H. wrote the manuscript.

\section*{Funding}
This work was funded by the Max Planck Society, the Natural Sciences and Engineering Research Council of Canada (P.H., R.J.D.M.), the Fonds de recherche du Qu\'ebec (P.H.), and the BWFGB Israel-Hamburg project LOM~2018 (R.B.).

\section*{Acknowledgments}
We acknowledge many helpful discussions with and support for modifying \emph{CrystFEL} by Thomas White.
We thank Anton Barty, Valerio Mariani, and Oleksandr Yefanov for making \emph{peakfinder8} available under the terms of the GNU Lesser General Public License.

\section*{Supplemental Data}
A set of example Jupyter notebooks explaining the processing pipeline in detail is included with this paper in PDF format.
The notebooks themselves, including all ancillary files required to reproduce our workflow can be downloaded at https://github.com/robertbuecker/serialed-examples.

\section*{Data and Code Availability Statement}
The raw diffraction data and indexed/integrated \inlinecode{stream} files which the examples in this paper have been derived from is available at EMPIAR (https://empiar.org) using the accession code EMPIAR-10542.
The diffractem software, along with installation instructions, is available under the terms of the GNU Lesser General Public License 2.1 or higher at https://github.com/robertbuecker/diffractem.
The electron-enabled version of CrystFEL 0.9.1 is available at https://stash.desy.de/projects/MPSDED/repos/crystfel under the terms of the GNU General Public License 3.0.
CrystFEL 0.10.0 will include the required features by default.

\bibliographystyle{alpha}
\bibliography{SerialED_processing}

\end{document}